\documentclass[conference]{IEEEtran}
\usepackage[T1]{fontenc}
\usepackage{amsmath}
\usepackage[cmintegrals]{newtxmath}
\usepackage{url}
\usepackage{soul,color,graphicx,float,epstopdf}
\usepackage{tablefootnote,textcomp,gensymb}
\usepackage{eurosym,booktabs,multirow}
\usepackage[caption=false]{subfig}
\usepackage{amsfonts} 
\usepackage{algorithmic}
\usepackage{array}
\usepackage{stfloats}
\usepackage{mathtools}
\usepackage{pdfpages}
\usepackage{comment}
\usepackage{balance}
\usepackage{xcolor}
\usepackage[normalem]{ulem}
\usepackage{enumitem}
\usepackage[implicit=false]{hyperref}
\usepackage{epstopdf}

\newcounter{tempEquationCounter} 
\newcounter{thisEquationNumber}

\definecolor{Orange}{rgb}{1,0.5,0}
\newcommand{\todo}[1]{\textsf{\textbf{\textcolor{Orange}{[#1]}}}}

\begin{document}
\title{Open RAN for 5G Supply Chain Diversification: The BEACON-5G Approach and Key Achievements}

\author{\IEEEauthorblockN{
Adnan Aijaz, Sajida Gufran, Tim Farnham, Sita Chintalapati, Adrián Sánchez-Mompó, Peizheng Li
}\\ 
\vspace{-2.00mm}
\IEEEauthorblockA{
\IEEEauthorrefmark{0} Bristol Research and Innovation Laboratory, Toshiba Europe Ltd., U.K.\\
Email: {\{firstname.lastname\}@toshiba-bril.com}
}}
\maketitle
\begin{abstract}
Open RAN brings multi-vendor diversity and interoperability to mobile/cellular networks. It is becoming part of governmental strategies for diversifying telecoms supply chains. This paper describes the approach and key achievements of the \emph{BEACON-5G} project, jointly funded by the UK government and industry. The BEACON-5G project aims at developing a competitive edge for 5G Open RAN and contributing toward its maturity. It addresses some of the key challenges in this respect and provides various innovations for system integration, network slicing, marketplace integration, cyber security, and white-box RAN. It also conducts real-world technology trials for urban use-cases. The paper also captures some of the key lessons learned during delivery, the main outcomes, and highlights potential impact on the wider UK 5G diversification strategy. 
\end{abstract}

\begin{IEEEkeywords}
5G, diversification, digital twin, marketplace, network slicing, O-RAN, RIC, system integration, xApps.
\end{IEEEkeywords}

\section{Introduction}
\vspace{-1.00mm}
\IEEEPARstart{C}{ommercialization} of 5G technology comes at an exciting time when the global mobile communications industry is witnessing a transformation, underpinned by openness and democratization, particularly in the radio access network (RAN). The broader focus of recent \emph{Open RAN} initiatives and frameworks is opening the protocols and interfaces between various components of the RAN and disaggregation which means running software stack on general-purpose vendor-neutral hardware. 

Telecommunications networks like 5G play an important role in almost every segment of modern-day life. These networks underpin critical national infrastructure; hence their security and resilience are crucial for delivering economic and social benefits. Recently, various governments worldwide have adopted the strategy of protecting the telecommunications infrastructure from high-risk vendors. In the UK, the government released its \emph{5G diversification strategy}\footnote{\url{https://www.gov.uk/government/publications/5g-supply-chain-diversification-strategy}} in December 2020. This strategy sets out an ambitious plan to diversify the telecoms supply chain. A key element of this 5G diversification strategy is to accelerate the development and deployment of open-interface 5G solutions to remove dependency on any single vendor and to create an open, innovative, resilient, and competitive market.  Open RAN solutions provide multi-vendor interoperability for MNOs and pave the way toward resilient and diversified supply chains and markets. 

In July 2021, the UK Department for Digital, Culture, Media and Sport (DCMS) launched the Future RAN competition (FRANC)\footnote{\url{https://www.gov.uk/guidance/future-ran-diversifying-the-5g-supply-chain}}, to allocate up to £30 million of R\&D funding to projects that support government’s 5G  diversification strategy. FRANC aimed at helping incentivize industry to create new products and services to unlock the full potential of Open RAN technology. BEACON-5G\footnote{\underline{B}uilding R\underline{E}configurable, \underline{A}gile, Se\underline{C}ure, and Trustw\underline{O}rthy Systems for Ope\underline{N}ness in \underline{5G}} is one of the winning FRANC projects. The BEACON-5G project provides various technological innovations toward the 5G diversification strategy. BEACON-5G's vision is to realize a high-performance multi-vendor 5G Open RAN system\footnote{In this paper, \emph{5G Open RAN system} refers to an end-to-end 5G system aligned with the O-RAN architecture~\cite{o2021ran}. } aligned with the Open RAN framework and principles, with native capabilities of openness, reconfigurability, service architecture agility, and resilience, for operation in dense urban local/private as well as public/carrier environments. 

To this end, this paper provides an overview of the BEACON-5G project, discussing its approach toward 5G diversification and highlighting the key achievements. The paper describes some of the key challenges (Section \ref{sect_approach}) addressed by the project for realizing its vision. It covers some of the key innovations (Section \ref{sect_innovations}) of the project toward multi-vendor system integration, end-to-end network slicing for reconfigurability, RAN digital twin for cyber security, marketplace integration for RAN democratization, interworking of white-box and vendor-based system components. It provides details of 5G Open RAN deployment in the field (Section \ref{sect_fd}). The paper also covers some of the lessons learned during delivery along with the key outcomes and the impact from the 5G diversification perspective (Section \ref{sect_impact}).    

\begin{figure}
\centering
\includegraphics[scale=0.3]{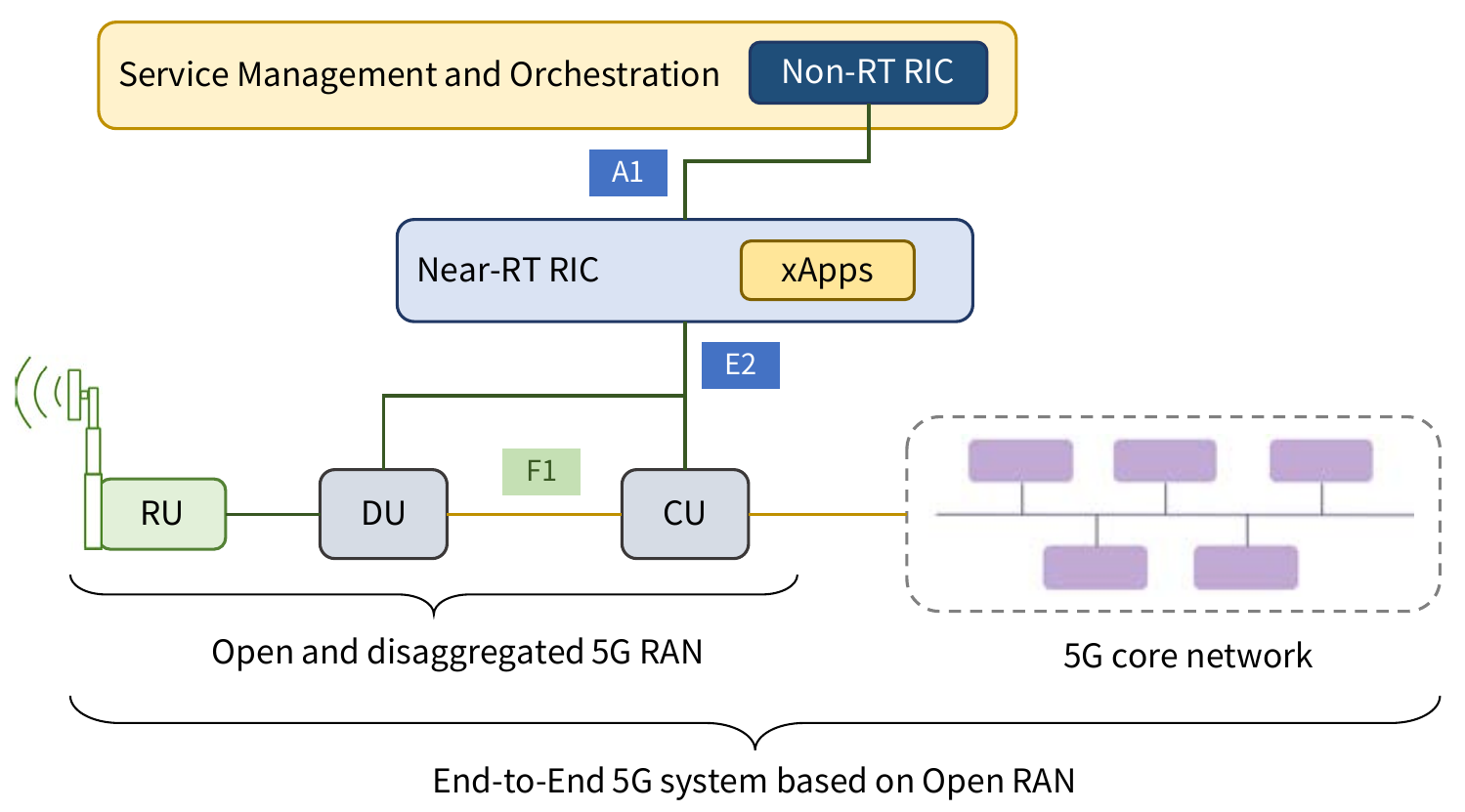}
\caption{5G Open RAN architecture (aligned with the O-RAN specifications).}
\label{arch_fig}
\vspace{-5.00mm}
\end{figure}

\section{Preliminaries on Open RAN}
\vspace{-1.00mm}

The Open RAN architecture is built on two key principles of openness and intelligence.  Openness revolves around the adoption of open and standardized interfaces for interoperability. This facilitates seamless integration of hardware components from different vendors, paving the way for multi-vendor RAN ecosystem. The O-RAN Alliance (\url{https://www.o-ran.org/}) has defined various specifications in this regard. On the other hand, the concept of intelligence revolves around the integration of artificial intelligence (AI) and machine learning (ML) capabilities in the RAN and enhanced network automation for resource control and management/orchestration~\cite{o2021ran_ai_ml}. Intelligence is underpinned by the newly-introduced architectural components, RAN intelligent controllers (RICs). Depending on the time sensitivity of tasks, applications can be residing at the non-real-time (non-RT)~\cite{o2021ran_Non_RT_ric} RIC or the near-real-time (near-RT) RIC~\cite{o2021ran_near_RT_ric}, also known as rApps or xApps, respectively.

Open RAN follows 3GPP specifications, e.g., the New Radio (NR) specifications for 5G, consisting of the centralized unit (CU), the distributed unit (DU), and the radio unit (RU). RU and DU are disaggregated based on the well-known 7.2x split, and connected via the open fronthaul interface. 
Further division of the CU yields two logical components, the CU control plane (CU-CP) and the user plane (CU-UP). 
The interconnection of the DU and CU transpires through the open midhaul F1 interface, branching into F1-C for control plane and F1-U for user plane connectivity. Establishing connectivity between the near-RT RIC and the CU/DU is achieved via E2 interface, facilitating near-RT control. The near-RT RIC connects with the non-RT RIC through the A1 interface which serves as the conduit for non-RT control and the implementation of AI/ML model updates within the near-RT RIC. The O1 interface links the non-RT RIC with other RAN components for management and orchestration. A more detailed overview of Open RAN  is given in~\cite{10024837}. Fig.~\ref{arch_fig} illustrates a representative example of 5G Open RAN system.

\section{BEACON-5G's Approach for 5G Diversification}\label{sect_approach}
\vspace{-1.00mm}
BEACON-5G's approach for 5G diversification revolves around providing a competitive edge for 5G Open RAN and contributing toward its maturity. The former is achieved by bringing conventional 5G capabilities like network slicing as well as through building new capabilities like marketplace integration, digital twins, and interworking with white-box RAN components. The latter is achieved through end-to-end system integration and technology trials based on real-world use-cases and deployments. 
Some of the key challenges addressed by the BEACON-5G project are discussed as follows. 

\subsection{Multi-vendor 5G System}\vspace{-1.00mm}
While Open RAN promises to break vendor lock-in, the viability of an end-to-end 5G system heavily relies on the successful interworking of multi-vendor system components, which requires compliance with interface specifications and interoperability testing. Moreover, such a system's performance is dictated by the performance of individual system components which may differ in hardware and software capabilities.

\subsection{Reconfigurability}\vspace{-1.00mm}
Reconfigurability empowers service providers to realize customized 5G networks over a common physical infrastructure while supporting diverse deployment scenarios and use-cases. It becomes particularly important for private 5G networks~\cite{Pvt_5G}, operating in either standalone or shared public-private mode, as well as for public 5G networks operating on the neutral host model. Reconfigurability requires end-to-end slicing capabilities with tight functional isolation among slices along with automation of management and orchestration capabilities. However, end-to-end slicing in 5G Open RAN systems becomes particularly challenging due to disaggregated and multi-vendor RAN components, new network elements and interfaces, and architectural limitations. 

\subsection{Open RAN Marketplace}
\begin{figure}[t]   
        \centering   
        \includegraphics[scale=0.11]{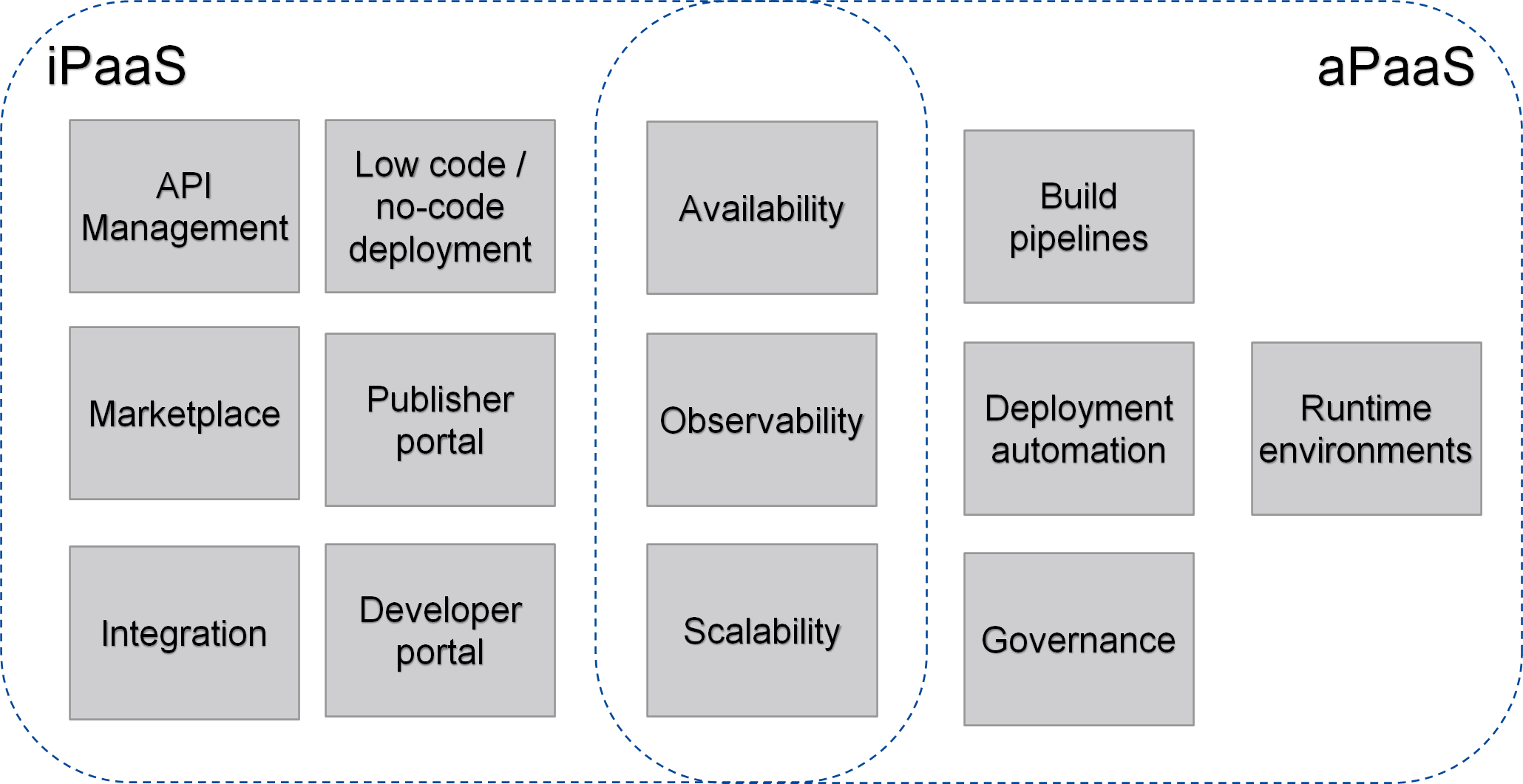}   
            \caption{Marketplace approach based on iPaaS and aPaaS in BEACON-5G.} \label{fig:architecture}
            \vspace{-5.00mm}
\end{figure}

\vspace{-1.00mm}
A marketplace serves as a means to realize collaborative revenue sharing, eliminating the need for intricate proprietary agreements or contracts between each participant. Certain current marketplace models concentrate on specific use-cases and services. 
For instance, NECOS~\cite{H2020-NECOS} offers Slice as a Service in marketplace, aligning resource requests across tenants and providers, whereas 5GTANGO~\cite{5GTANGO} provides a  marketplace with a customizable orchestrator, slice manager, and service mapper.
The BEACON-5G marketplace strategy emphasizes software integration across diverse deployment settings, utilizing the API-centric integration Platform-as-a-Service (iPaas) and application PaaS (aPaaS) model aligned with Open RAN standards, as shown in Fig.~\ref{fig:architecture}. This approach takes advantage of common and standardized APIs to facilitate an open eco-system, enabling tailored deployment and intelligent control~\cite{10.3389/frcmn.2023.1127039}. 
 
\subsection{Digital Twin}\vspace{-1.00mm}
Digital twins have gained significant attention within the realm of commercial 5G networks, because the complexity of scaling 5G networks and the stringent performance requirements pose challenges for the overall deployment process of 5G. Functioning as a virtual replica of the actual network, digital twins provide a valuable platform for cost-effective and immediate algorithmic development and verification. 
However, in many cases, digital twin is treated merely as an external simulator, and the correlation/interplay between the digital twin and the real network have only been loosely explored until now. This approach limits the effectiveness of the digital twin in guiding 5G network operations due to elongated feedback loops and detachment from real-world systems. Moreover, utilizing an external digital twin introduces higher latency and brings unneglectable risks of information leakage.

\subsection{Cyber Security}\vspace{-1.00mm}
The Open RAN architecture is characterized by decoupled hardware and software, and introduces new interfaces and components which broaden the threat/attack surface of the network. The resilience of open interfaces and general-purpose hardware against cyber security threats/attacks is mostly unknown. Similarly, the trustworthiness of 5G Open RAN systems for critical and safety-of-life applications in industrial and consumer domains must be investigated, especially from the perspective of building additional capabilities (e.g., privacy/security assurance). 

\subsection{Open RAN Trials}\vspace{-1.00mm}
Real-world technology trials of 5G Open RAN systems are still in infancy, particularly trials in urban environments focusing on key capabilities, interoperability, and end-to-end performance, and new use-cases.

\subsection{White-box RAN}\vspace{-1.00mm}
The open innovation ecosystem around 5G is attractive for developing white-box 5G RAN system components which are based on general-purpose hardware and open-source software stack. The components pave the way for WiFi-like diversified 5G supply chains, provide transparency and promote economies of scale. However, the interworking of such white-box RAN components with vendor-based system components and their performance aspects remain largely unexplored.

\section{BEACON-5G: Key Innovations}\label{sect_innovations}
BEACON-5G provides various innovations for addressing the aforementioned challenges.

\subsection{5G Open RAN System Integration}

\begin{figure}
\centering
\includegraphics[scale=0.32]{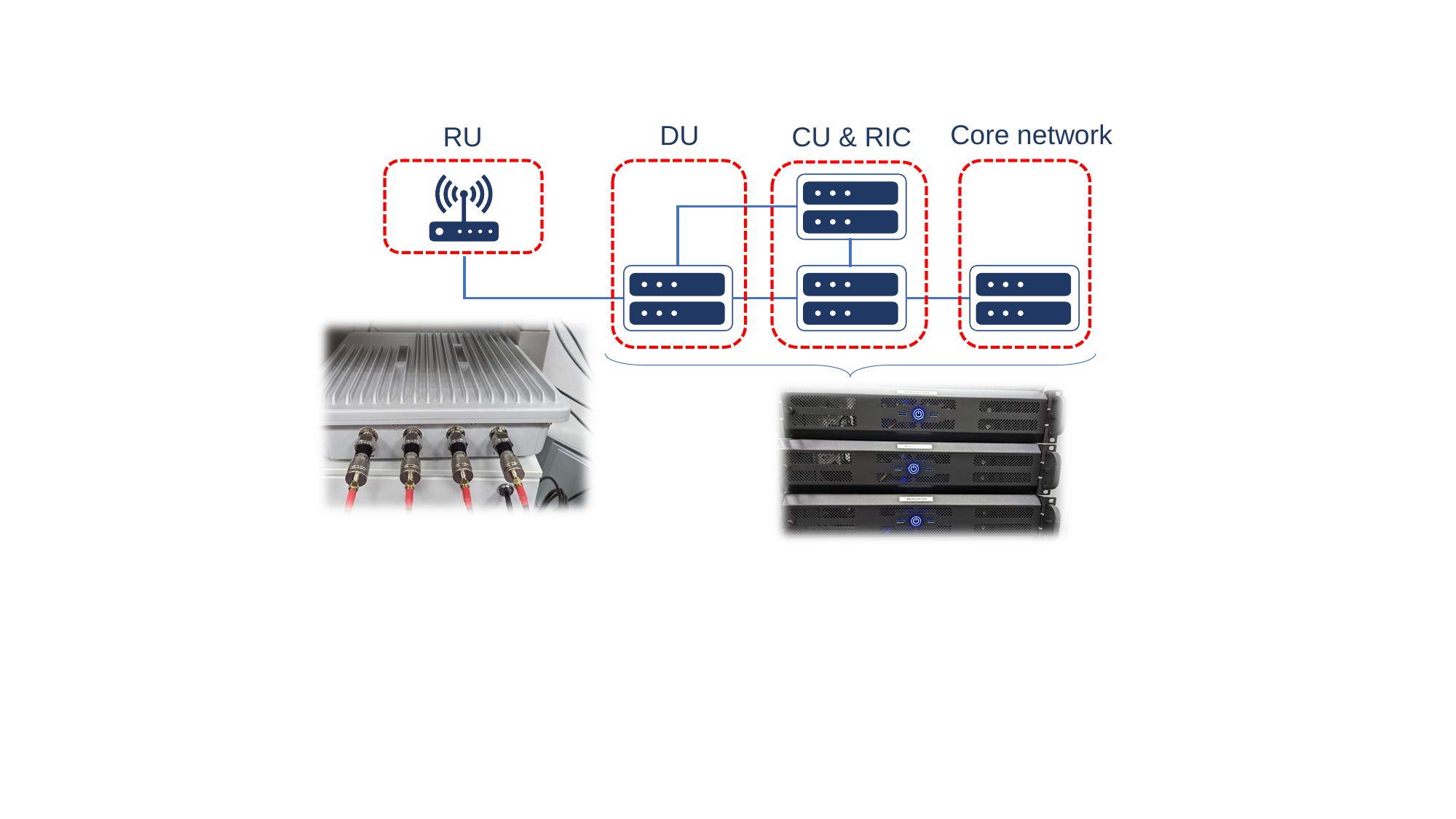} 
\caption{Illustration of the end-to-end 5G Open RAN system. }
\label{5G_syst}
\vspace{-2em}
\end{figure}

\begin{figure}
\centering
\includegraphics[scale=0.23]{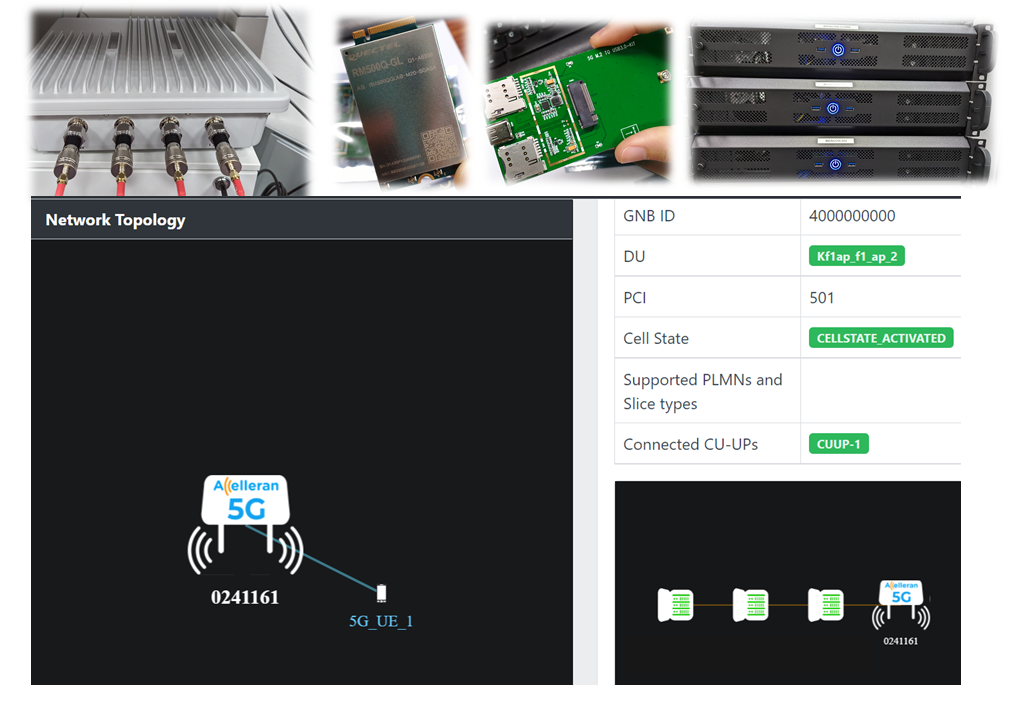} 
\caption{RAN dashboard showing connectivity status.}
\label{sys_int}
\vspace{-2em}
\end{figure}

The BEACON-5G project carried out system integration of an end-to-end multi-vendor 5G standalone system (compliant with 3GPP Release 15), which is illustrated in Fig.~\ref{5G_syst}. The end-to-end system includes three general-purpose high-specification servers, running 5G DU, CU, RIC stacks, and 5G core network stack, wherein merely the RU is the proprietary hardware component. The suppliers of this system are listed as follows. 
\begin{itemize}
    \item \textbf{RU} - Benetel (outdoor unit; RAN650\footnote{\url{https://benetel.com/ran650/}})
    \item \textbf{DU} - Effnet/Phluido and IS-Wireless stacks
    \item \textbf{CU} - Accelleran stack
    \item \textbf{RIC} - Accelleran stack
    \item \textbf{Core network} - Attocore stack
\end{itemize}

The BEACON-5G system operates in band n77u (3.8 - 4.2 GHz) which is the private 5G band in the UK, while the Quectel RM500Q-GL module is adopted as the primary 5G device. BEACON-5G system provides dashboards for the RAN and the core network for configuration and management. Fig.~\ref{sys_int} shows a live cell and successfully connected UE in the RAN dashboard. 

\begin{figure}
\centering
\includegraphics[scale=0.43]{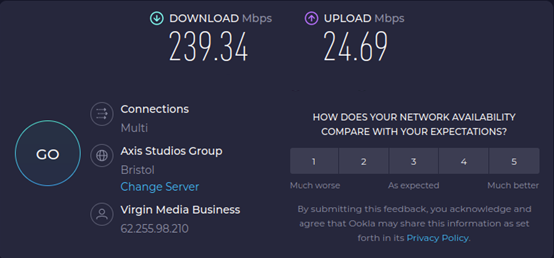} \ 
\includegraphics[scale=0.349]{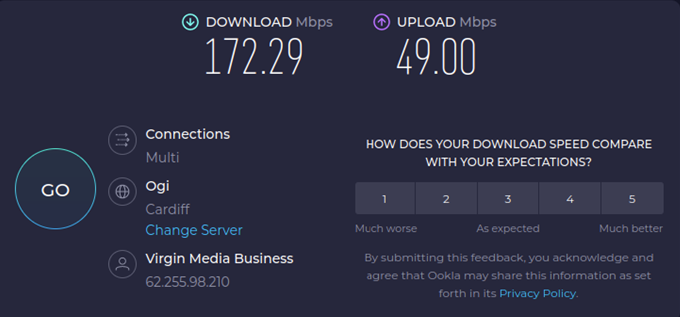}
\caption{Throughput performance of the 5G Open RAN system.}
\label{Thpt_perf_fig1}
\end{figure}

\begin{figure}
\centering
\includegraphics[scale=0.18]{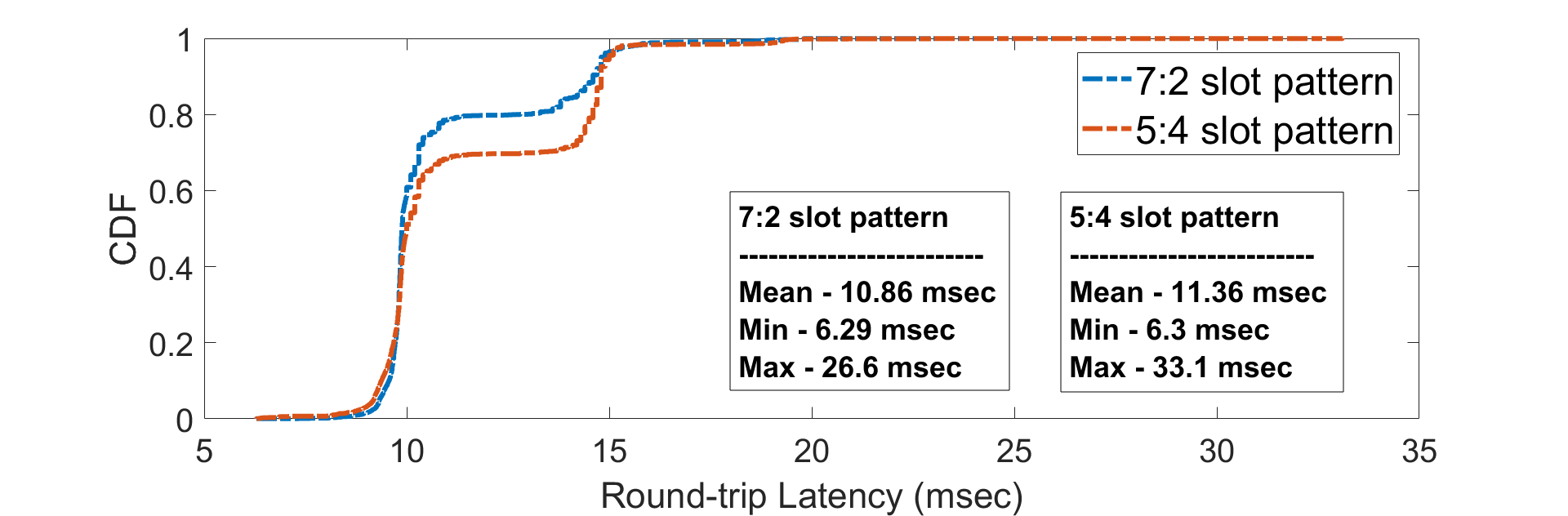} 
\caption{Latency performance of the 5G Open RAN system. }
\label{latency_perf}
\vspace{-1.5em}
\end{figure}

The downlink and uplink throughput performance of the BEACON-5G system with 40 MHz bandwidth is demonstrated in Fig.~\ref{Thpt_perf_fig1}. The DU supports different configurations for supporting different services. For instance, in the downlink-centric configuration, where 7 slots (and 6 symbols) are allocated for downlink and 2 slots (and 4 symbols) are allocated for uplink, the achieved downlink and uplink throughput is approximately 240 Mbps and 25 Mbps, respectively. In contrast, in the uplink-centric configuration, where 5 slots (and 6 symbols) are allocated for the downlink and 4 slots (and 4 symbols) are allocated for the uplink, the achieved downlink and uplink throughput is 172 Mbps and 49 Mbps, respectively. The testing with 100 MHz bandwidth is also conducted in BEACON-5G, which achieves a peak downlink throughput of more than 500 Mbps in the downlink-centric configuration. The latency performance of the system is shown in Fig. \ref{latency_perf}. The average latency is approximately 10 msec which is also the requirement for most real-time control applications. 

\begin{figure}
\centering
\includegraphics[scale=0.26]{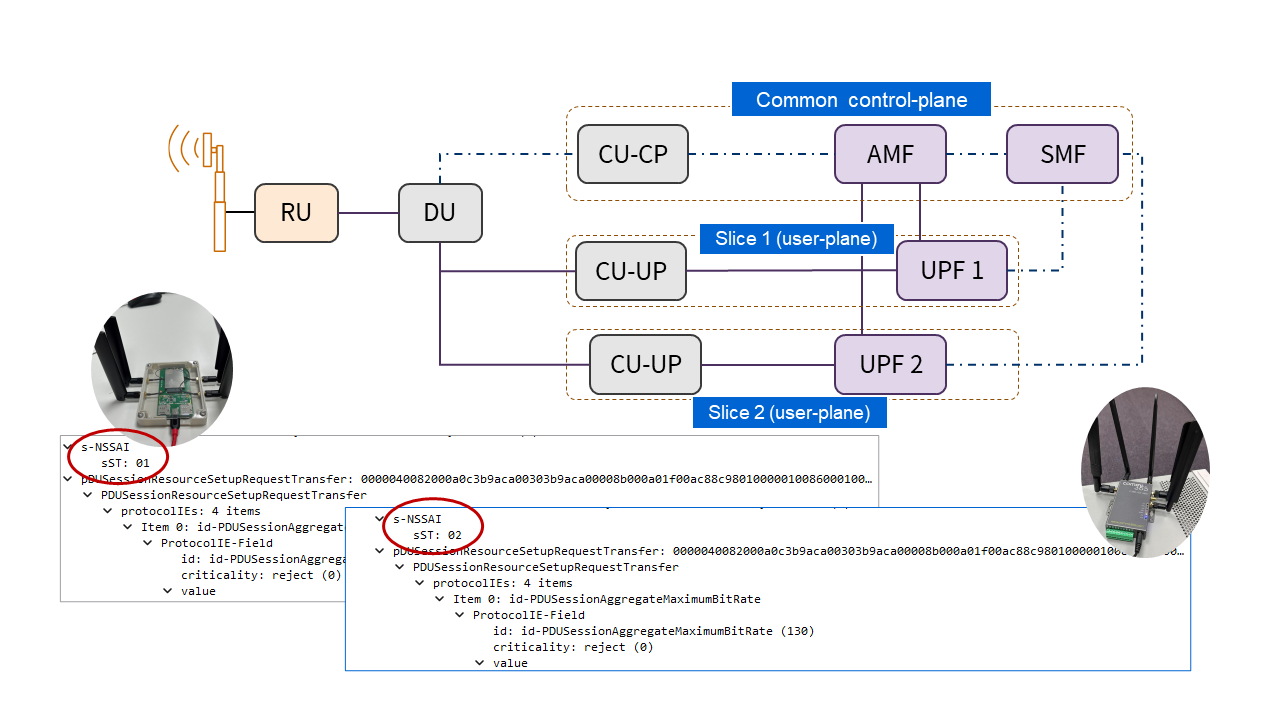}
\caption{End-to-End network slicing and slice validation.}
\label{slicing_fig}
\vspace{-1.5em}
\end{figure}

\subsection{End-to-End Network Slicing}
BEACON-5G addresses the reconfigurability challenge through an end-to-end network slicing solution. The key principles of BEACON-5G's slicing approach are as follows. 

\begin{itemize}
    \item Exploit the programmability offered by O-RAN architecture for network slicing. 
    
\item Modular and dynamic slice composition extending to different network components in the 
RAN and the core network. 

 \item Provide functional isolation among slices for strict performance guarantees. 
 
\item Automation of network slice management and orchestration functions. 
\end{itemize}

\begin{figure}
\centering
\includegraphics[scale=0.2]{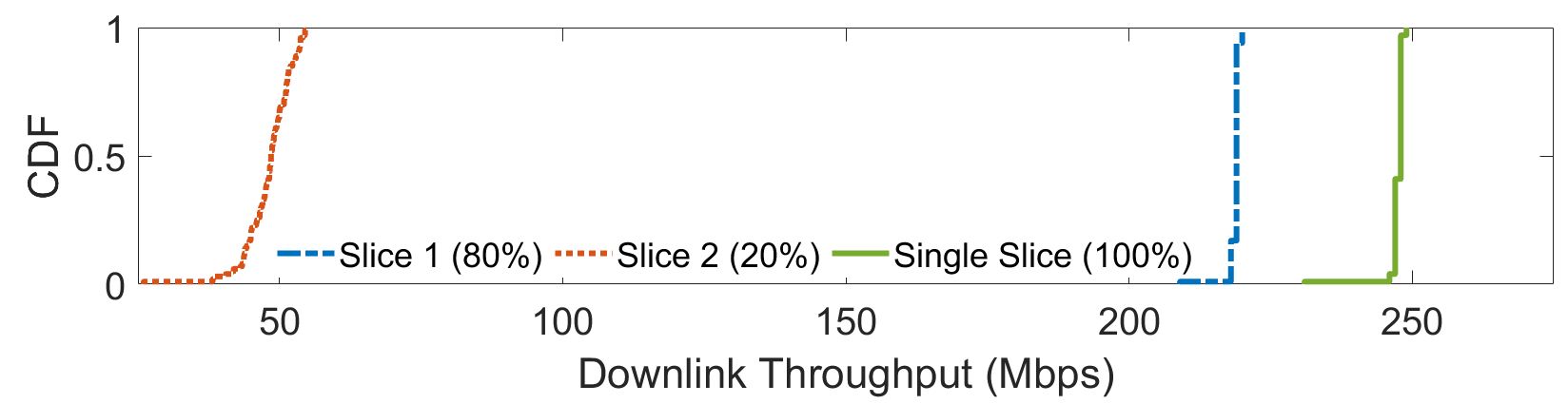} 
\caption{Performance of radio resource slicing at the DU.}
\label{radio_slicing_perf}
\end{figure}

BEACON-5G has developed an end-to-end network slicing solution with RIC-centric management and orchestration of slices. A slice consists of subnets in the RAN and the core network. Slicing in the RAN consists of radio slicing in the DU and slice-specific deployment of CU, i.e., control-plane instance (CU-CP) and  user-plane instance (CU-UP). Both DU and CU provide slice-specific telemetry information to the near-RT RIC for slice management.  Slicing in the core network consists of slice-specific network functions including the user-plane function (UPF), the session management function (SMF), and the access and mobility management function (AMF). 
The RIC handles real-time management of slices and deployment of network functions in the RAN as well as in the core network. Fig. \ref{slicing_fig} shows a slicing scenario where there is a common control-plane and dedicated user-planes for two slices. The two UEs are connected on two different slices.

We have two DU stacks one of which supports slicing of radio resources. The performance of radio resource slicing functionality is shown in Fig. \ref{radio_slicing_perf}. With single slice, the average downlink UDP throughput for a UE is 244.9 Mbps. With two slices, i.e., Slice 1 and Slice 2, allocated 80\% and 20\% of the total radio resources, the average downlink throughput is 216.64 Mbps and 47.72 Mbps, respectively.


\subsection{5G RAN Digital Twin for Anomaly Detection}
\begin{figure}
\centering
\includegraphics[scale=0.65]{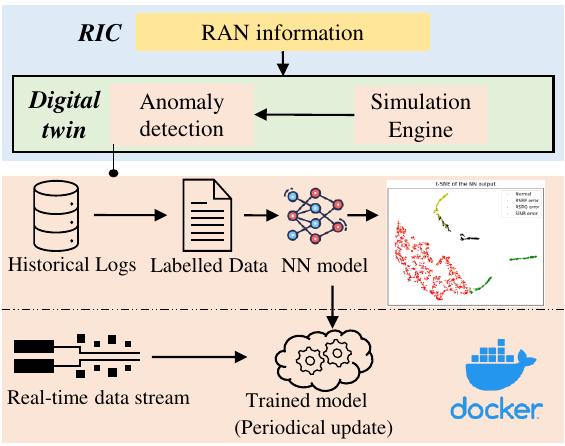}
\caption{RAN digital twin functionalities and anomaly detection workflow.}
\label{fig:dt workflow}
\vspace{-1.5em}
\end{figure}

BEACON-5G introduces an innovative approach to designing and implementing a digital twin, which harnesses the programmable capabilities of the RIC, integrating the digital twin into the RIC as an integral component of RAN operations. One of the use-cases of the RAN digital twin is centered around cyber security for detecting anomalies in network connectivity. On the technical front, the digital twin is constructed using a simulation engine/module, an anomaly detection module, real-time UE and system information, along with historical logs from various RAN components. This approach yields three key advantages: Firstly, this digital twin enables near real-time system performance simulation (with latency at 10ms level). Secondly, it facilitates on-board closed-loop inference and anomaly detection, thereby enhancing responsiveness and data security. Thirdly, the digital twin can engage with other RIC applications to facilitate information exchange, fusion, and policy control. An instance of UE connectivity anomaly detection utilizing this digital twin methodology is presented in Fig.~\ref{fig:dt workflow}. This digital twin captures real-time operational information and fed it to the two modules for optimal scheduling and monitoring UE-level connectivity. The anomaly detection module is constructed on the dockerized NN model consting the completed training, test, deployment and updating loop. More details of this implementation are elaborated on~\cite{li2023demo}.

\subsection{Marketplace Integration}
\begin{figure}[t]   
        \centering   
        \includegraphics[scale=0.11]{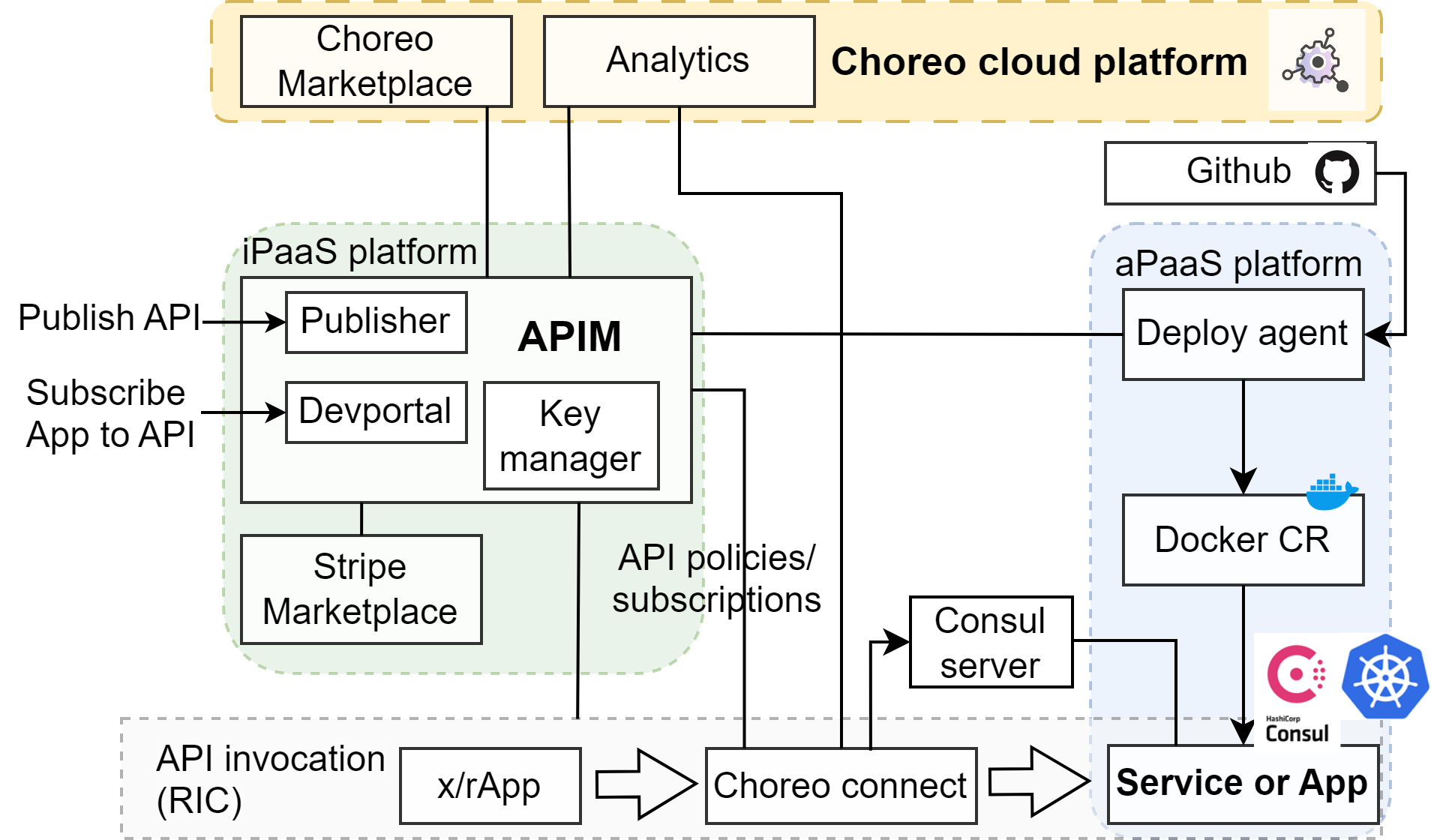}   
            \caption{Marketplace integration architecture within BEACON-5G.} \label{fig:integration}
            \vspace{-5.00mm}
\end{figure}

The marketplace integration is realized through the use of WSO2-based iPaaS components together with vendor specific deploy agents to various aPaaS environments (as shown in Fig.~\ref{fig:integration}). 
The main components of the iPaaS are the API manager, and plugins supporting Stripe marketplace billing for reconciliation. The API manager supports the publisher and developer portals for publishing and subscribing to APIs, respectively. In this manner, the monetization occurs through business models or plans associated with the APIs, such as subscriptions with quotas, pay per use, or performance-related service level agreement. The aPaaS environments abstract the physical environment resources through use of Kubernetes and federated Consul service meshes, which utilize flexible deployment scripts that are shared and accessed through GitHub. The final platform component is the Choreo cloud platform which supports the centralized marketplace, with external connectors to external services and applications. Further details of the BEACON-5G marketplace can be found in~\cite{Farnham2023demo}.

\subsection{Interworking of White-box RAN and Vendor Components}

In the context of Open RAN, a white-box component is defined as off-the-shelf standard hardware/software systems that can be used for the deployment of O-RAN-compliant software stacks, all the relevant interworking in these white-box components can be viewed, are well-known, and understood by the Open RAN community. With interoperability one of the main challenges of the open RAN movement, it is only logical to try and leverage open-source solutions as a testing ground for interoperability with both vendor and open-source components.

Activities on white-box  interworking of white-box system components in BEACON-5G  rely on the use of OpenAirInterface (OAI)-based stack as open-source RAN providers. The potential of OAI stack in technology trials has been shown in various studies, e.g., \cite{Open_5G_Netsoft}. As a part of this we have verified (i) the OAI monolithic gNB with vendor-based core and (ii) split CU/DU with vendor-based core. A comparison of them is shown in Table~\ref{table:OAI_performance}.

\begin{table}[ht]
\centering
\caption{Performance of White-box RAN with vendor core}
\label{table:OAI_performance}
\begin{tabular}{ccc}
\toprule
OAI version               & OAI CU-DU split & Monolithic OAI gNB    \\ \midrule
Round trip latency (avg.) & 43.35ms     & 16-17ms  \\ 
Iperf UL & 4-6Mbps & 2Mbps \\
Iperf DL & 5-10Mbps & 120Mbps\\
\bottomrule
\end{tabular}
\end{table}

Further,  we conducted assessments to confirm the compatibility of different RANs and near-RT RICs, encompassing both open-source and vendor-based systems. We employed  open-source RIC from the O-RAN software community (ORAN-SC) and tested the maturity of the E2 interface (connecting the near-RT RIC and the CU/DU). With both ORAN-SC RIC (open-source) or Vendor RICs , E2 setup is successful when we are using vendor DU. But the message IE’s are empty. There is no info going in these IEs. So it seems only the initial level of handshake is handled as of now. When we are using OAI DU (open-source), the initial E2 setup request is seen to be sent by both RICs but no reply comes after from DU. It appears from the logs, OAI DU is not decoding the E2 setup message correctly.

\subsection{Technology Trials}
BEACON-5G conducted both horizontal and vertical technology trials to show the potential of 5G Open RAN system and the key innovations of the project in real-world environments. The project conducted two verticals-centric trial, including one for the smart city and the other for remote assistance use-case. It also conducted horizontal trials focusing on interoperability and cyber security. Due to page limitations, we have not included the details of all the technology trials. However, the smart city trial is discussed in Section~\ref{sect_fd}.




\section{Field Deployment}\label{sect_fd}
\begin{figure}
\centering
\includegraphics[scale=0.3]{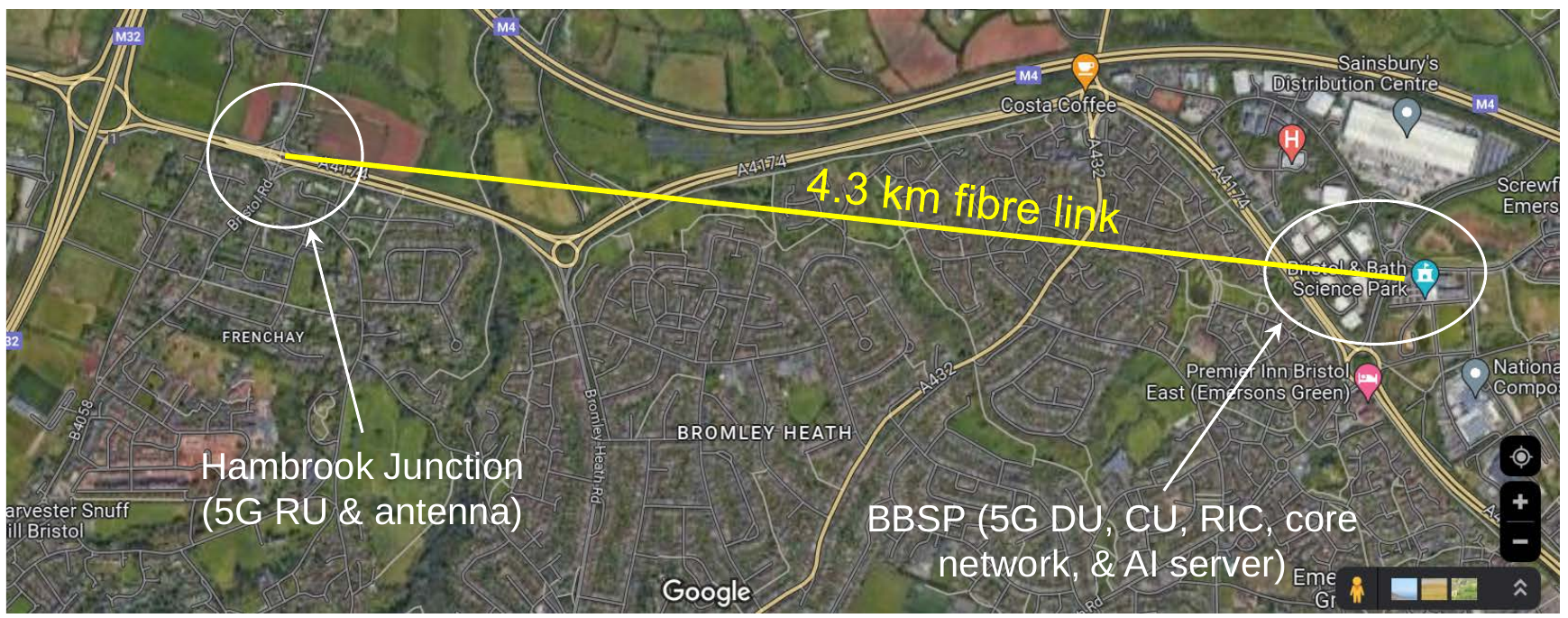}
\caption{Location of the BEACON-5G smart city use-case trial deployment.}
\label{deploy_fig}
\end{figure}

As part of BEACON-5G technology trials, the multi-vendor 5G Open RAN system has been deployed in the field. The deployment is spread across two sites, as shown in Fig. \ref{deploy_fig}, and it is for the smart city use-case (discussed later). Open RAN reduces the footprint of 5G deployment at the junction such that only the RU and the antenna are deployed at the junction (with necessary power and connectivity infrastructure) while other system components are located at another site. The two sites are connected by a 4.3 km fiber link.

\subsection{Smart City Use-case}
5G technology is promising to improve various aspects of urban life including public safety and security, 
transportation and mobility improvements, infrastructure and environmental monitoring, and managing large-scale sensor networks. Together with AI, edge-computing, and Internet-of-Things (IoT) technologies, it empowers local authorities to modernize core technology 
infrastructure and to improve citizen services through digital transformation of different sectors.

This use-case explores the benefits brought by 5G technology in general and Open RAN technology in particular for the smart cities vertical. The main focus of this use-case is dynamic urban traffic 
control using low latency and high data rate capabilities of 5G technology and enhanced support for running AI models in Open RAN systems.

Dynamic urban traffic control is achieved through an edge application that resides at the near-RT RIC. One or more wirelessly-connected cameras (e.g., CCTVs) 
capture the traffics at the junction and the video feed is transmitted to the RIC. The edge application, running AI algorithms, utilizes real-time video information and 
provides an optimized timing plan for junction control which ensures a smooth traffic flow. The timing 
plan is sent to the traffic junction via the 5G network.


\subsection{ML Architecture Overview}

\begin{figure}
\centering
\includegraphics[scale=0.33]{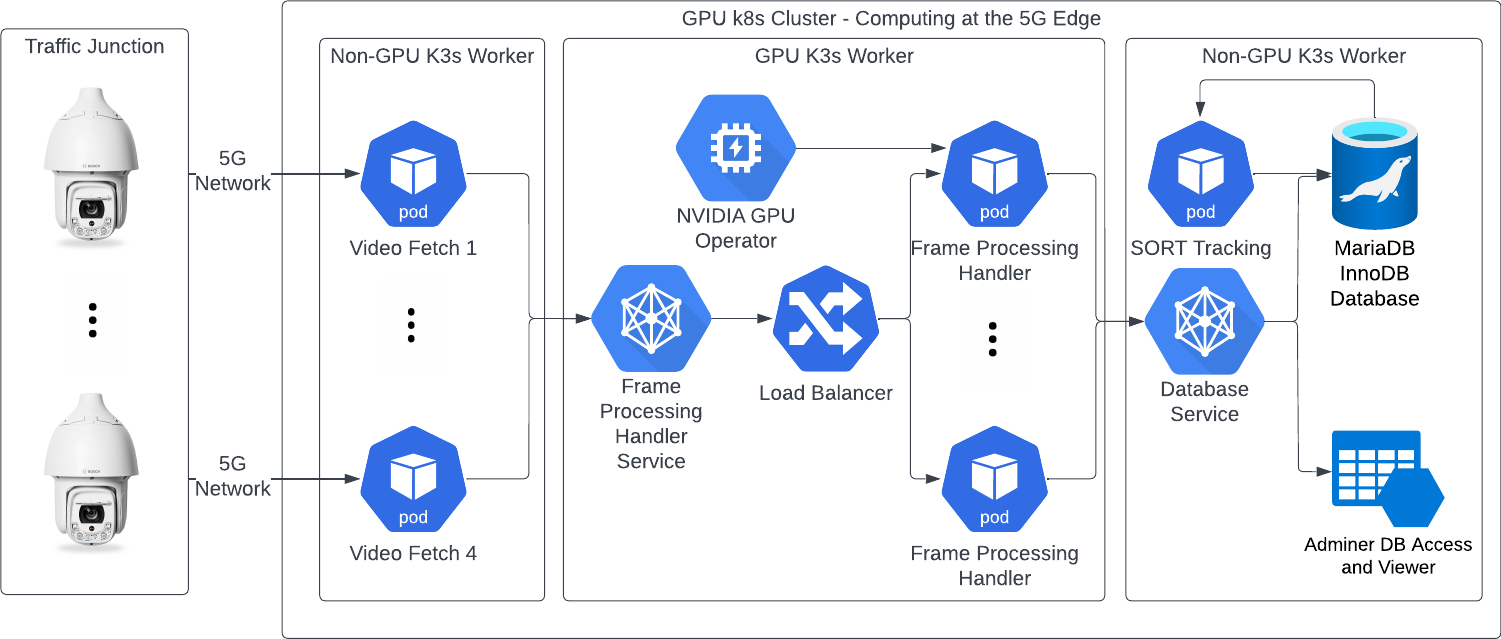}
\caption{ML processing pipeline.}
\label{ML_Processing_Pipeline}
\end{figure}

Our advanced ML pipeline processes image data from four cameras at the junction directly at the edge-core, ensuring speed and efficiency. The architecture comprises three main components: Video Fetch Pods, Frame Processing Handlers, and a Central Database, as can be seen in Fig. \ref{ML_Processing_Pipeline}. The Video Fetch Pods retrieve RTSP streams from the cameras, converting them into frames. These frames are then sent to the Frame Processing Handler Pods via a Load Balancer. These handlers generate bounding boxes for vehicles, classify them, and after Non-Maxima Suppression, store the results in the database.
This edge-centric architecture reduces data transmission and improves efficiency. We've transitioned from a CPU-based model with a latency of one second to a GPU-based version, aiming for 10-20 milliseconds latency, harnessing the full potential of the 5G network for real-time analytics.

\subsection{Vehicle Detection and Classification via Bounding-Box}

\begin{figure}
\centering
\includegraphics[scale=0.34]{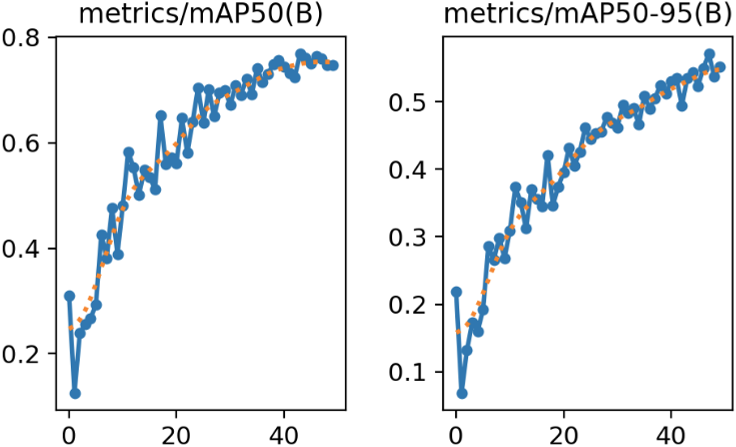}
\caption{mAP on IoU of 50\% and 90\% of the bounding box proposals.}
\label{mAP50}
 \vspace{-1.5em}
\end{figure}

Leveraging a five-category vehicle detection bounding-box dataset ~\cite{Dataset_OID_v4_Vehicles} based on the Open Images Dataset v4~\cite{OID4}, we used the YOLOv8-m model \cite{Jocher_YOLO_by_Ultralytics_2023} to suggest bounding boxes and categorize vehicles into cars, motorcycles, trucks, buses, and bicycles.
As a foundational starting point, we adopted a pre-trained model that learned on the MS COCO dataset, targeting bounding box proposal and classification tasks. After fine-tuning it on our dataset for 50 epochs, the model achieved a decent Mean Average Precision (mAP) for bounding boxes with an Intersection over Union (IoU) of 50\% or more, coming close to 0.8, as shown in Fig. \ref{mAP50}.

\begin{figure}
\centering
\includegraphics[scale=0.35]{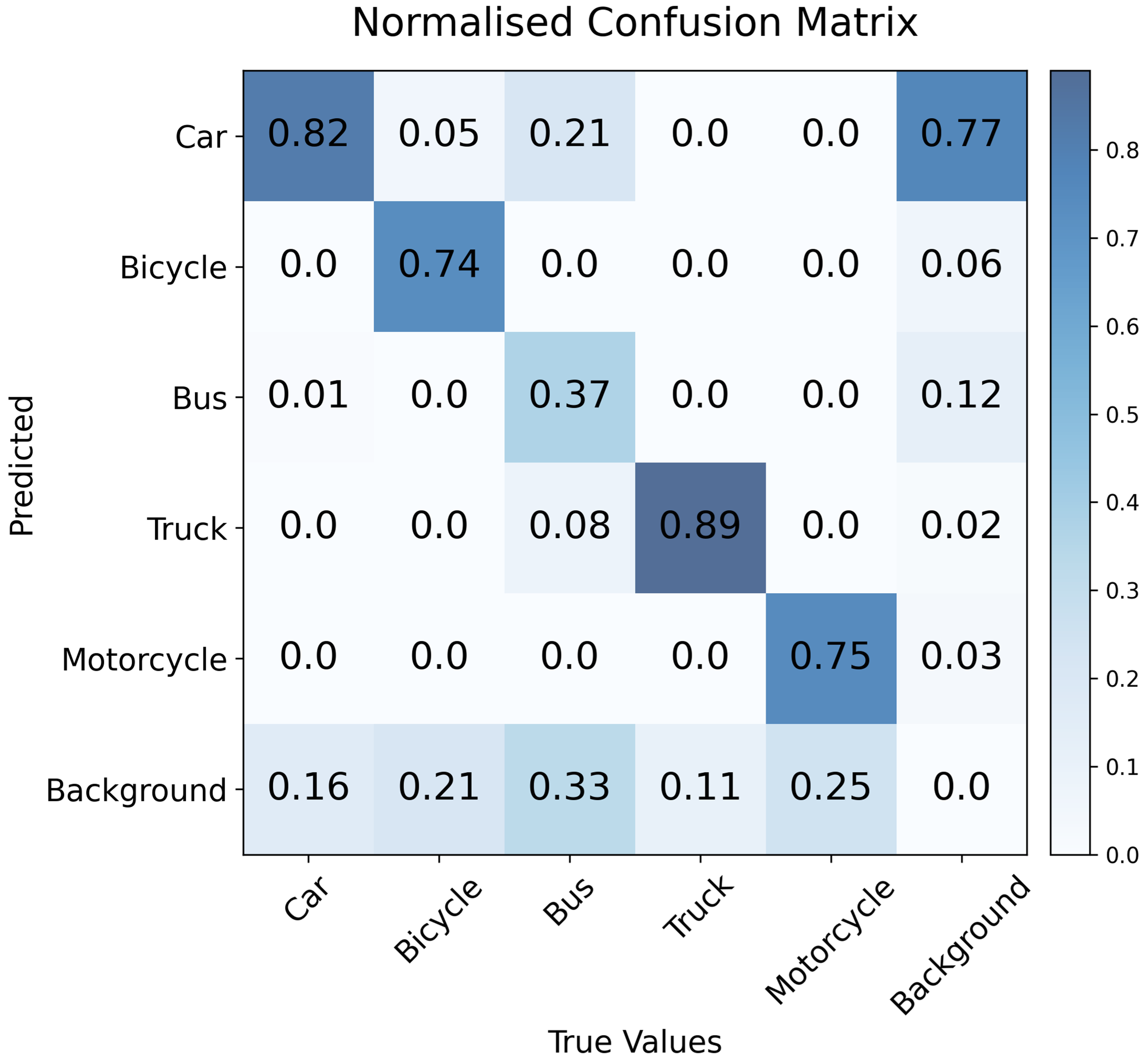}
\caption{Normalised confusion matrix for the 5 classes and background.}
\label{confusion_matrix}
\end{figure}

While these results are promising, there's room for improvement. Using a smaller dataset and a compact neural network meant that some metrics, like Precision, Recall, and mAP, didn't reach top-tier scores. This impacted the confusion matrix as seen in Fig. \ref{confusion_matrix}, especially with less represented categories like buses and trucks that can look quite similar. But these findings give us a good starting point for future improvements.

\subsection{Operational Performance Post-Deployment}

Following the deployment of our trained model, a discernible drop in accuracy was observed, an anticipated outcome considering the geographic specificity of the original dataset and the temporal discrepancy with our application scenario. This manifested predominantly as a heightened misclassification between Trucks and Buses, more marked than those observed in the test data.

Regarding latency, the model exhibited an average processing time of 11.8 ms. This duration encompasses the image pre-processing, vehicle detection, vehicle classification, and non-maxima suppression of the proposed bounding boxes.

\begin{figure}
\centering
\includegraphics[scale=0.4]{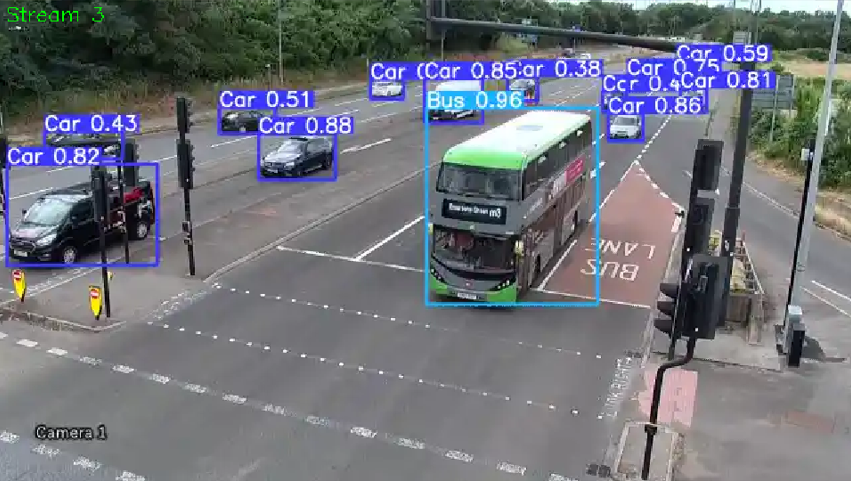}
\caption{Example of frame with detections, confidence and class overlayed from one of the four cameras of the Smart City deployment.}
\label{camera_stream}
\vspace{-1.5em}
\end{figure}

The resulting stream with the detection overlays from the YOLOv8 framework provide the classes and the confidence of the model with the bounding box candidates over the original image on real time as seen in Fig. \ref{camera_stream}.

\section{Lessons, Outcomes, and Overall Impact}\label{sect_impact}
Some of the key lessons learned during the BEACON-5G project are summarized below. 

\begin{itemize}
    \item The Open RAN ecosystem is still evolving and system components with different capabilities exist, for example, DUs with or without radio slicing functionality, which can create various challenges for system integrators and operators. 
    \item Deployment of 5G in the field is simplified with devices supporting power-over-Ethernet (PoE). 
    \item Compliance with 3GPP and O-RAN specifications and quality of implementation are crucial factors dictating swappability of different system components. 
\end{itemize}

Some of the key outcomes of the  project and potential impact on 5G diversification is highlighted as follows. 

\begin{itemize}
    \item The system integration activity has demonstrated the technological viability of open-interface technologies and end-to-end multi-vendor 5G systems. This not only brings new vendors in the 5G supply chain but also new players in the value chain. 
    
    \item The development activities exploited the flexibility and programmability offered by the O-RAN architecture and led to solutions for network automation, resource management/orchestration and anomaly detection. Such RIC-centric solutions show the potential for third-party solutions/services for the RAN, ultimately paving the way toward RAN democratization of the RAN.

    \item The field trial for smart city use-case demonstrates the capabilities of 5G Open RAN for local/private networks for city authorities helping in their digital transformation objectives and contributing toward increased penetration of 5G technology. 

    \item The smart city technology trial leverages AI techniques running at the edge which can be tightly integrated in the RAN for network automation and real-time control-centric applications \cite{edge_AI}. 

    \item The interworking activities provide enhanced insights into the maturity of different system components and interfaces. 

    \item Overall activities also demonstrate the proof of business viability as system integrator and service/solution provider for 5G Open RAN systems. 
\end{itemize}


\section{Concluding Remarks}
Open RAN is the key enabler for 5G supply chain diversification. The BEACON-5G project addressed crucial challenges around integration, enhancement, and maturity of 5G Open RAN systems. The key innovations of the project include multi-vendor system integration, end-to-end network slicing, digital twin for anomaly detection, marketplace integration, and interworking of white-box and vendor-based system components. Some of these innovations have been demonstrated through real-world deployment. The BEACON-5G project not only showed the technological viability of multi-vendor 5G systems but also the feasibility of operating as system integrator, private network operator, and services/solutions provider for 5G Open RAN systems, meeting some of the key objectives of the 5G diversification strategy.


\section*{Acknowledgements}
Successful system integration, technology development, and field deployment 
would not have been possible without efforts and support of many individuals including Lode Spruyt and Mahmoud Esawi from  Accelleran, Komal Sharma, Ashok Soundararajan,  
Ajith Sahadevan, Vianney Anis, Ben Holden, Ioannis Mavromatis, Adina Nistor, Yichao Jin, and Mahesh Sooriyabandara from Toshiba,  Jason Cooper and Marcin Kuzmicki 
from Attocore, Nita Patel, Paul Worseley, Ryan Brown, Dave 
Adams, Andrew Porter, and Michael Davenport from South Gloucestershire Council, and Brian Constable and Kevin 
Tyrrell from Select Electric.

The funding of FRANC projects by the UK government (DCMS) is gratefully acknowledged as well. 

\bibliographystyle{IEEEtran} %
\bibliography{IEEEabrv,references} 

\end{document}